# Cophasing the Planet Formation Imager

Romain G. Petrov[*a], Abdelkarim Boskri[b], Thami Elhalkouj[b], John Monnier[c], Michael Ireland[d], Stefan Kraus[e]

[a]Université de la Côte d'Azur, Laboratoire Lagrange, UMR7293, Bd. de l'Observatoire, 06304 Nice, France; [b]Université Cadi-Ayyad, LPHEA, Marrakech, Maroc; [c]University of Michigan, USA ; [d]University of Sydney, Australia, [e]University of Exeter, UK.

## ABSTRACT

The Planet Formation Imager (PFI) is a project for a very large optical interferometer intended to obtain images of the planet formation process at scales as small as the Hill sphere of giant exoplanets. Its main science instruments will work in the thermal infrared but it will be cophased in the near infrared, where it requires also some capacity for scientific imaging. PFI imaging and resolution specifications imply an array of 12 to 20 apertures and baselines up to a few kilometers cophased at near infrared coherent magnitudes as large as 10. This paper discusses various cophasing architectures and the corresponding minimum diameter of individual apertures, which is the dominant element of PFI cost estimates. From a global analysis of the possible combinations of pairwise fringe sensors, we show that conventional approaches used in current interferometers imply the use of prohibitively large telescopes and we indicate the innovative strategies that would allow building PFI with affordable apertures smaller than 2 m in diameter. The approach with the best potential appears to be Hierarchical Fringe Tracking based on "two beams spatial filters" that cophase pairs of neighboring telescopes with all the efficiency of a two telescopes fringe tracker and transmit most of the flux as if it was produced by an unique single mode aperture to cophase pairs of pairs and then pairs of groups of apertures. We consider also the adaptation to PFI of more conventional approaches such as a combination of GRAVITY like fringe trackers or single or multiple chains of 2T fringe trackers.

## 1. INTRODUCTION

The direct detection and characterization of earth twins is likely to be a very difficult objective that could not be reached in a close future, particularly if such a planet does not orbit a relatively nearby star. An alternative approach to find the probability of existence of any kind of planets, including earth twins, is to understand in detail the Planet Formation process. Checking evolution scenarios requires many high-resolution images of planetary formation disks at different ages and for a range of stellar masses. This is the main science goal the of the very large optical interferometer project called PFI for "Planet Formation Imager".

The resolution and the contrast must be sufficient to investigate the interaction between giant planets and the protoplanetary disk, down to the Hill sphere of giant planets. This sets PFI main specifications:

- Main science operation: N band.
    - Secondary science operation for the main science goal of Planet Formation Imaging: K and L bands
    - Science goal for PFI: Q band
    - Secondary science goal: AGNs in the visible, stellar physics, etc.
- Cophasing in the near infrared to allow coherent integration in the N band.
- Kilometric baselines: to achieve 0.1 au at 100 pc and at 10 µm, we need $B_{max}$>2 Km.

[*] romain.petrov@unice.fr; phone +33 4 92 00 39 61, +33 4 92 07 64 08.

- Number of apertures: $n_T$>12; goal $n_T$>20. In this paper we use the intermediate value $n_T$=16.
- Limiting coherent magnitude: $H_C$=10. We need H>9 (or K>9) to cover a sufficient sample of targets. In the H band the flux for most targets will be dominated by the central source, with a typical core angular size ~0.13 mas for H~10, i.e. a visibility for the maximum baseline $V_*(B_{max})$~0.45. This yields the coherent magnitude specification H~10.

In optical interferometry, the fringes are stabilized by a device called the Fringe Tracker (FT), which measures and corrects in real time the variations of Optical Path Difference (OPD) introduced by the atmosphere and by interferometer instabilities. Stabilizing the fringes within a small fraction of wavelength is called "cophasing" and is based on a "phase delay" measurement. Maintaining the fringes in the center of the coherence length $R\lambda$ (where $R$ is the instrument spectral resolution) is called "coherencing" and is based on a "group delay" measurement.

So far, most FTs are either pair-wise or all-in-one.

- Pair-wise, like PIONIER[1] and GRAVITY[2], which means that the differential piston (the average OPD between two apertures) is measured for *all* possible telescope pairs. This is the system most often used for cophasing.
- All-in-one, like the instruments AMBER[3] and MATISSE[4], which means that *all* apertures produce a common global interferogram where the different baselines are separated by the data processing because they have different spatial frequencies. This system is mostly used when only a coherencing is desired.

In both cases, the performance of the system decreases with the number of apertures and that implies that the minimum size of the individual telescope necessary to achieve a given sensitivity increases. As PFI needs more apertures than any existing optical interferometer, the conventional pair-wise or all-in-one approaches are likely to need very large telescopes. For example, the best current fringe tracker is quite certainly the one installed in the 2$^{nd}$ generation VLTI instrument GRAVITY. It's a 4 telescopes "all pairs" pair-wise system that use state of the art integrated optics and an avalanche photo diodes detector with a read-out noise lower than 1e$^-$. Its first results on the sky show that it achieves limiting magnitudes of K~7.5 with the 4 AT (1.8m diameter) and K~10.5 with the 4 UTs (8 m diameter). With much more than 4 telescopes, the required diameter necessary to achieve K or H ~10 is therefore certainly larger than 8 m with a system based on the GRAVITY design. That would make PFI unaffordable. Thus we have to consider alternative FT strategies.

It has been shown that the fundamental performances of the pair-wise and all-in-one approaches are similar, with a small advantage for the pairwise approach. In the following we will therefore consider only variations based on pair-wise combinations.

## 2. ACCURACY AND SPECIFICATIONS OF A FRINGE TRACKER

A fringe tracker has two fundamental functions. The main one is to measure the rapidly variable phase of the fringes (the phase delay) and stabilize it. If this is executed on nearly monochromatic fringes, there is a $2\pi$ ambiguity on the measurement and therefore a $\lambda$ ambiguity on the correction. Thus, phase delay measurement allows undetected fringe jumps. A fringe jump in the K band is a fraction of a fringe shift in the N band that will make coherent fringe integration impossible. Thus, the second function of a fringe tracker is to prevent fringe jumps or at the very least to detect them. This is performed by an analysis of the group delay of the whole fringe packet. In current systems, this is obtained by dispersing the light over $n_\lambda$ spectral channels. Then, the global $\lambda$ ambiguity is removed until the smallest common multiple of all channels wavelength. Coherencing is also a key feature for the robustness of the FT operation as it speeds up the acquisition and re-acquisition of fringes. Thus, the two fundamental parameters of a combination of pair-wise fringe trackers are:

- The number $n_{pair}$ of pair-wise FT fed by each aperture.
- The number $n_\lambda$ of spectral channels used for coherencing.

### 2.1 Accuracy of a phase delay measurement

The accuracy $\sigma_{\varphi 1}$ of a phase delay measurement in a single spectral channel is given by[5] $\sigma_{\varphi 1}=1/(\sqrt{2}\ SNR_{C1})$ where $SNR_{C1}$ is the signal-to-noise ratio on the coherent flux in one spectral channel. If we consider only the source photon noise and the detector readout noise and neglect the thermal background in the near infrared to compute[6] the coherent flux SNR, we get:

$$\sigma_{\varphi 1} = \frac{1}{\sqrt{2}SNR_{C1}} \simeq \frac{\sqrt{2\frac{n_*}{n_\lambda n_{pair}}+n_{pix}\sigma_{RON}^2}}{\sqrt{2}V\frac{n_*}{n_\lambda n_{pair}}} \qquad (1)$$

If the phase accuracy per spectral channel is smaller than about 1 radian[5] ($\sigma_{\varphi 1}<1$), the measurement from all spectral channels can be combined to get the global phase delay accuracy $\sigma_\varphi$:

$$\sigma_\varphi = \frac{\sigma_{\varphi 1}}{\sqrt{n_\lambda}} \simeq \frac{\sqrt{2\frac{n_*}{n_\lambda n_{pair}}+n_{pix}\sigma_{RON}^2}}{\sqrt{2}V\frac{n_*}{n_\lambda n_{pair}}\sqrt{n_\lambda}} \qquad (2)$$

where

- $n_*$ is the number of coherent photons received from the source from each aperture.
- $n_{pix}$ is the number of pixels used for the measurement. Pairwise setups generally use the so-called "ABCD" approach that can be fairly approximated by equation (2) with $n_{pix}=4$.
- $\sigma_{RON}$ is the standard deviation of the detector readout noise.
- $V$ is the instrument visibility (the source visibility affects the coherent flux $n_*$).

## 2.2 Specifications of a fringe sensor

A fringe sensor must satisfy three conditions.

- The phase delay measurement in each spectral channel must be more accurate than 1 radian: $\sigma_{\varphi 1}<1$. For larger uncertainties, the combination of several spectral channels becomes very inefficient[7].
- The piston measurement accuracy $\sigma_p = \lambda\sigma_\varphi/2\pi$ must be better than a given specification $\lambda/p_{spec}$. In the following we shall use $p_{spec}=10$. This yields:

$$\sigma_{\varphi 1} = \sigma_\varphi\sqrt{n_\lambda} < \frac{2\pi\sqrt{n_\lambda}}{p_{spec}}$$

- The group delay accuracy $\sigma_G$ must be small with regard to the coherence length. Typically $\sigma_G < R\lambda/10$. If the phase delay is measured in a sufficient number $n_\lambda$ of spectral channels, this conditions is obtained when the first condition ($\sigma_{\varphi 1}<1$) is satisfied.

These conditions can be combined in

$$\sigma_{\varphi 1} < min\left\{1, \frac{2\pi\sqrt{n_\lambda}}{p_{spec}}\right\} \qquad (3)$$

Combining equations (2) and (3) and solving the 2nd degree equations yields the limiting coherent flux

$$n_{*min} = n_{pair}.max\left\{n_\lambda \frac{1+\sqrt{1+2n_{pix}\sigma_{RON}^2 V^2}}{2V^2}, \frac{1+\sqrt{1+2n_\lambda n_{pix}\sigma_{RON}^2\left(\frac{2\pi}{p_{spec}}V\right)^2}}{2\left(\frac{2\pi}{p_{spec}}V\right)^2}\right\} \qquad (4)$$

The number of photons collected from the source by each aperture is given by:

$$n_* = n_{0m}.10^{-0.4m}\frac{\pi}{4}D^2.\delta\lambda.T.S_{tr}.\tau \qquad (5)$$

where:

- $n_{0m}$ is the number of photons received outside the earth atmosphere from a source of magnitude m=0 per unit of surface, time and spectral bandpass. For example $n_{0V}=1084\ 10^4$ photons/cm$^2$.s.µm in the visible, $n_{0H}=58\ 10^4$ photons/cm$^2$.s.µm in the H band and $n_{0K}=43\ 10^4$ photons/cm$^2$.s.µm in the K band.

- $m$ is the magnitude of the source in the band used for cophasing.
- $D$ is the telescope diameter, in meters.
- $\delta\lambda$ is the overall spectral bandpass used for cophasing, in microns.
- $T$ is the global transmission, from sky to computer.
- $S_{tr}$ is the telescope Strehl ratio.
- $\tau$ is the exposure time used for each phase delay measurement.

Combining equations (4) and (5) yields the minimum telescope diameter required to operate the fringe tracker and achieve the fringe tracking efficiency $\lambda/p_{spec}$ at the required magnitude $m_{spec}$.

$$D_{min}^2 = \frac{10^{0.4 m_{spec}}}{0.79\, n_{0m}.\delta\lambda.T.S_{tr}.\tau} n_{pair} \cdot max\left\{ n_\lambda \frac{1+\sqrt{1+2 n_{pix}\sigma_{RON}^2 V^2}}{2V^2}\ ,\ \frac{1+\sqrt{1+2 n_\lambda n_{pix}\sigma_{RON}^2 \left(\frac{2\pi}{p_{spec}}V\right)^2}}{2\left(\frac{2\pi}{p_{spec}}V\right)^2}\right\} \quad (6)$$

For a modest tracking accuracy ($p_{spec}<10$) and standard number of spectral channels ($n_\lambda \geq 3$), the first term in {} dominates and we have:

$$D_{min}^2 = \frac{n_{pair} n_\lambda}{T.\tau}\frac{10^{0.4 m_{spec}}}{0.79\, n_{0m}.\delta\lambda.S_{tr}}.\frac{1+\sqrt{1+2 n_{pix}\sigma_{RON}^2 V^2}}{2V^2} \quad (7)$$

The parameters $m_{spec}$ and $\delta\lambda$ are specifications. The typical value for a pair wise fringe tracker is $n_{pix}=4$. The readout noise $\sigma_{RON}=1$ is set by state of the art detectors and a high instrumental visibility $V>0.9$ can be achieved on single mode fast sampled instruments. These values can hardly be changed. The design effort must therefore be concentrated on decreasing $n_{pair}$ and $n_\lambda$ and increasing $T$ and $\tau$.

## 3. CRITICAL PARAMETERS FOR PFI COPHASING ARCHITECTURE

Let us discuss more in details the parameters that define the cophasing architecture of PFI.

### 3.1 Spectral band and bandpass used for cophasing

Cophasing will be performed in the near infrared where we can combine excellent detectors with $\sigma_{RON}^2 < 1$ with seeing perturbations much smaller than in the visible. We have considered using the full near infrared range from the J to the K bands (1 to 2.45 mm), but, on one side, our typical sources are much fainter in J than in H and K and on the other side, K band is expected to provide important scientific information about the dust and gas interaction for example. So we have reserved for the fringe tracker the H band, centered at 1.65 mm and with a width δλ=0.35 mm. It might be worth remembering that, if this results critical, we could use also all or a part of the K band for cophasing.

### 3.2 Fringe tracker accuracy

As the main science instrument is at larger wavelength, a modest specification of $p_{spec}=10$ (i.e. $\lambda/10$) in the H band is sufficient.

### 3.3 Limiting sensitivity

As explained in the introduction, the desired PFI limiting sensitivity is the coherent magnitude $H_C=10$.

### 3.4 Instrument visibility

All considered systems use single mode filtering on each beam and fast time sampling. Thus we can assume an instrumental visibility $V=0.9$, which is actually achieved in the GRAVITY FT. The source visibility loss is included in our specification based coherent magnitude.

### 3.5 Strehl ratio from the adaptive optics on each aperture

We assume that each individual aperture has an Adaptive Optics correction used to feed the interferometer with single mode beams both for science and for cophasing. The performance of the AO system depends on the telescope aperture, the wavelength and the cost of the system. The AO systems operational or in development for the VLTI telescopes show that a Strehl ratio of 50% can be obtained in the near infrared for telescopes ranging from 1.8 m to 8 m in diameter with intermediate cost AO systems for sources up to H>10. Therefore, in the following we always assume the conservative $S_{tr}=0.5$. However, a full study of the performances and cost of affordable AO systems for PFI remains an open issue.

### 3.6 Global transmission

We use a global transmission, from sky to computer, of *T=1%* as in the actual GRAVITY Fringe Tracker. This is often considered as a typical value for the VLTI in the near infrared but it is in fact quite pessimistic. The global transmission for AMBER in the K band is of the order of 4%. A careful design of the interferometer can certainly allow substantially higher transmissions, with a direct and strong impact on the cost of PFI. A gain of a factor 4 in transmission allows a decrease of individual telescope diameter by a factor 2 and a gain in telescope cost much larger than 4.

### 3.7 Fringe sensor exposure time

We set the fringe sensor exposure time to $\tau = 5\ ms$. This is quite pessimistic with regard to the coherence time of the atmospheric piston. For example, Folcher et al[8] show that the optimum exposure time, near the limiting magnitude and in standard Paranal seeing, should be between 10 and 20 ms. However, experience shows that $\tau = 5\ ms$ is actually optimistic at Paranal with the UTs mainly because of the telescope vibrations. The design of PFI should include a priori the reduction of vibrations in the interferometer, either by mechanical design or by including from the beginning an active sensing and correction of the fast piston perturbations introduced by the interferometer itself.

### 3.8 Number of spectral channels in the fringe sensor

The phase delay can be measured with a single spectral channel but the kλ ambiguity results in fringe jumps, loss of fringes and a slow fringe acquisition and reacquisition procedure. A group delay measure is therefore necessary. This is generally obtained by dispersing the fringe sensor light in $n_\lambda$ spectral channels. A typical number is $n_\lambda=5$, as in the GRAVITY fringe tracker. Mourard et al[9] have shown that the optimum value for this approach is $n_\lambda=3$. Some authors[10] have proposed to use $n_\lambda=1$, with the possibility to switch rapidly to a "coherencing mode" with $n_\lambda \geq 3$ but the operational efficiency of such an approach remains unknown. In the Hierarchical Fringe Tracker (HFT)[11], which is briefly described below, we use $n_\lambda=1$, because the group delay measurement is obtained from a different device with the photons that are transmitted by the "pair-wise" levels. Optimizing $n_\lambda$ is a key parameter of the design of the PFI fringe tracker and a key motivations for the HFT concept.

### 3.9 Number of cophasing pairs fed by each aperture

In the standard "pair-wise" approach, the flux of each aperture is divided between $n_{pair}=n_T-1$ pairs. Equations (6) and (7) show that this is extremely costly in terms of aperture size requirements. We have therefore to consider architectures where each telescope feeds directly a smaller number of cophasing pairs. Figure 1 shows the minimum "chain configuration" where each telescope is cophased only with its two neighbors and $n_{pair}=2$. This might be well suited for a ring like interferometric array, in particular to observe resolved objects where only the shortest baselines yield a high coherent flux. A probable drawback is a strong sensitivity to the loss of fringes in one fringe tracker. Then the piston for all longer baselines will be affected by the addition of errors in many individual FTs. To relax this we could use more complex chains, as the one shown in figure 2 where each aperture is cophased with 4 other ones ($n_{pair}=4$). This introduces some redundancy and thus safety in the reconstruction of the piston on all short baselines. An alternative way to use a $n_{pair}=4$ configuration is displayed in figure 3, where each telescope feeds a local "GRAVITY like FT" plus an additional global one. This offers the advantage to use a well-evaluated device. Finally, figure 4 displays a setup based on our hierarchical fringe tracker concept[11]. Each pair of neighboring telescopes feeds a 2 beams spatial filter (SF2B) that transmits most (typically 75%) of the flux when the input beams are cophased as if it was produced by an unique telescope feeding a single mode beam. When the incoming beams are not cophased, up to 75% of the light is deflected to measure the piston difference. There are several ways to design such a SF2B. One could imagine an integrated optics "ABCD" system where two outputs (BC) are actually merged and propagated to the next level. The potential performance of such a device should be very close to this of a standard pairwise device cophasing two apertures: $\underline{n_{pair}=1}$. Then the outputs of the first level SF2B are used to cophase pairs of pairs, then pairs of quadruplets, etc. The flux used to cophase deeper levels increases, which could compensate for the loss of coherence due to the use of longer baselines in the pair of pairs and then pairs of groups. The flux transmitted through all levels, including the deeper one, can feed a group delay sensor, which removes the need to disperse in several spectral channels in the phase delay sensors: $\underline{n_\lambda=1}$. We are currently evaluating a HFT concept based on achromatic bulk optics SF2Bs. The computer simulations are very encouraging and a prototype is under development. If this concept works it would be optimum for the cophasing large interferometers such as PFI as it yields $n_{pair}=1$ and $n_\lambda=1$.

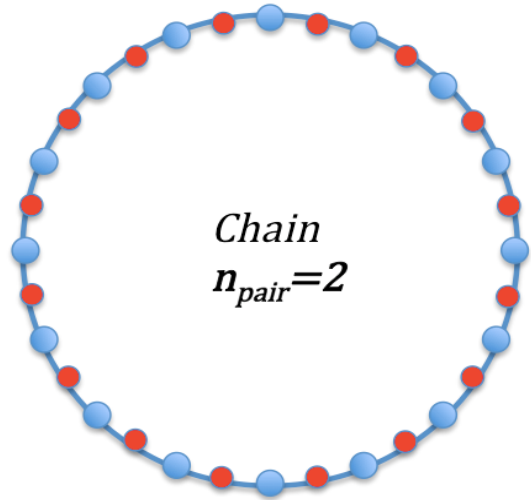

Figure 1: Simple Chain Setup. Each blue point represents a telescope and each red point a pair-wise FT. Each aperture is cophased only with its two neighbors, and its flux is shared between two pairs. This setup can be well adapted to ring interferometer architectures, particularly if the coherent flux on longer baselines is faint on extended sources. The drawback is the strong accumulation of noise on the OPD of longer baselines, in particular if the chain is broken because one FT lost tracking.

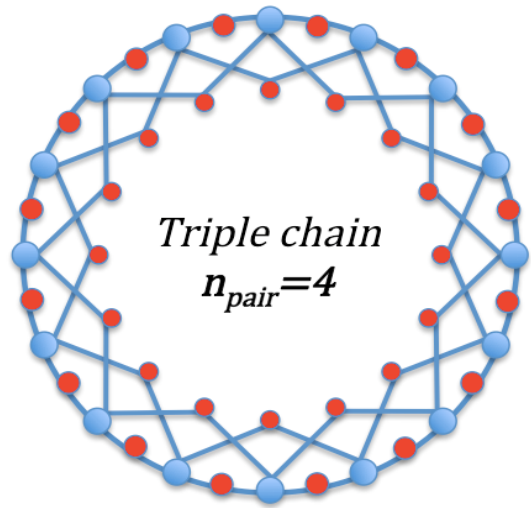

Figure 2: Triple Chain Setup. Each aperture if cophased with 4 other telescopes and participates to 3 FT chains. This setup will be much less sensitive than the simple chain to loss of fringes by some FTs, without being as costly in flux than the full pairwise setup.

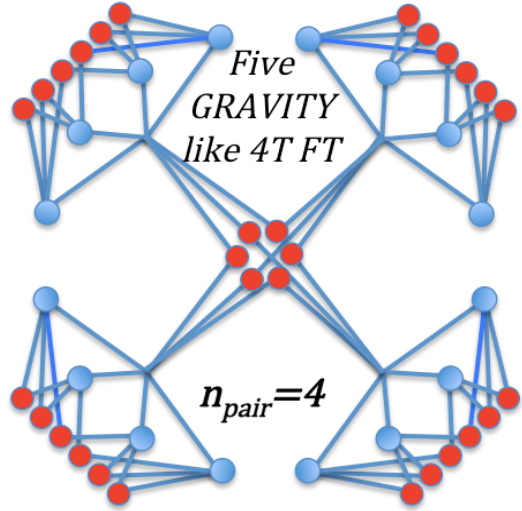

Figure 3: This setup is based on well tested "GRAVITY like" FTs. Each telescopes feeds a local "GRAVITY like" 4 beams FT and an additional global 4 beams tracker. The beams from the local group of 4 apertures are grouped than split in 3 to feed the global FT. This is a $n_{pair}=4$ configuration.

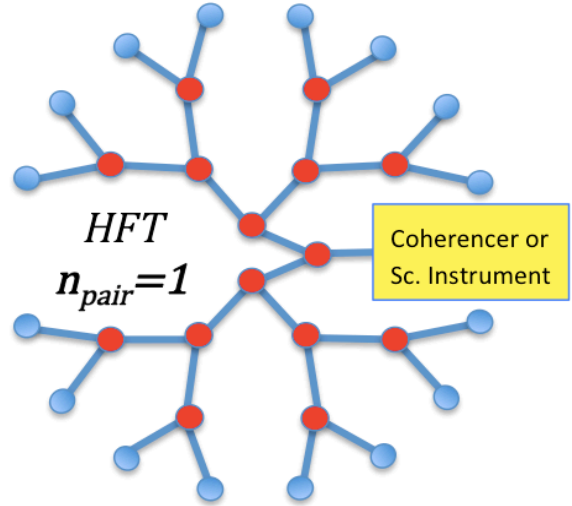

Figure 4: Hierarchical Fringe Tracking setup. We cophase pairs of apertures, then pairs of pairs, then pairs of quadruplets, etc… Each individual FT transmits most of the incoming flux when the pair is cophased as if it results from a single aperture. The flux transmitted by the deepest FT can be used in a scientific instrument or in a group delay sensor. The sensitivity of the cophasing is set by the first level of FT that use all the flux from a pair of telescopes: it is a $n_{pair}=1$ configuration. On the lower levels we have longer baselines but more flux in each FT.

### 3.10 Minimum diameter for PFI apertures

In the H band, with an overall transmission of $T=1\%$ and a sensor exposure time $\tau = 5\ ms$, equation (7) yields:

$$D_{min}^2 = 1.33\ n_{pair} n_\lambda\ m^2 \tag{8}$$

Applying equation (6) (which is equivalent to (7) and hence (8) for all values but $n_\lambda=1$) gives the following minimum dimater values to cophase a coherent magnitude $H_C=10$ source with an accuracy $\lambda/10$:

| Table 1: minimum diameter (in m) of individual PFI apertures $(T=1\%, \tau=5ms, S_{tr}=0.5, V=0.9, \delta\lambda=0.35~\mu m, \lambda/10$ accuracy in the H band) | | | | | |
|---|---|---|---|---|---|
| $n_{pair}$ | 1 | 2 | 3 | 4 | 15 |
| $n_\lambda=1$ | 1.2 | 1.7 | 2.1 | 2.4 | 4.6 |
| $n_\lambda=3$ | 2.0 | 2.8 | 3.5 | 4.0 | 7.7 |
| $n_\lambda=5$ | 2.6 | 3.6 | 4.5 | 5.2 | 10.0 |

## 4. CONCLUSION

We see that full pairwise solutions ($n_{pair}=15$, $n_\lambda=3$ or $n_\lambda=5$) are prohibitive. Optimized chains with $n_{pair}=3$ or $n_{pair}=4$ and $n_\lambda=3$ with conventional "ABCD" fringe sensors yields telescope diameters in the 3.5 to 4 m range. However this diameter could be reduced by a factor two if a careful design of the interferometer allows an overall transmission larger than 4% instead of the 1% achieved with GRAVITY on the VLTI. Unconventional solutions such as the HFT ($n_{pair}=1$, $n_\lambda=1$ or $n_\lambda=2$) would allow building PFI with apertures quite smaller than 2 m, in particular with a good transmission. The feasibility of these promising new approaches must therefore be investigated in detail. All configuration proposed here will show some sensitivity to the loss of fringes in local FTs. This must be investigated in terms of u-v coverage and ultimately image reconstruction efficiency. The global conclusion remains that there are serious tracks for research and development in cophasing toward a PFI with relatively small and therefore cheap apertures.